# Analysis Paralysis: When to stop?

Er.Akshay Bhardwaj,
University Institute of Information Technology,Himachal Pradesh University,Shimla-5(India)

**Abstract**

**Analysis of a system constitutes the most important aspect of the systems development life cycle.But it is also the most confusing and time consuming of all the stages.The critical question always remains: How much and till when to analyse? Ed Yourdon has called this phenomenon as Analysis Paralysis. In this paper, I suggest a model which can actually help in arriving at a satisfactory answer to this problem.**

**Table of Contents**

**1  Introduction**

There are seven major stages in the Systems Development Life Cycle.They are 1)Recognition of need 2)Requirements Analysis 3) SystemsAnalysis 4)Systems Design 5)System Testing 6)System Implementation 7)Post Implementation.Out of all these steps though all are equally important the most important among these is Systems Analysis. This stage by definition states: Systems Analysis is the dissection of a system into its component pieces to study how its component pieces interact and work.[1]. Broadly systems analysis phase encompasses the following main activities or sub phases: 1)Survey 2)study and 3)define.[2].These sub phases actually encompass the following activities.In 1)An effort is undertaken to study the existing system where the change to be brought is desired.In 2) the phase is studied and the techniques to bring about the desired change are examined.In 3) the new system is defined keeping in view the information that has been accumulated from the previous phases.However in all the above mentioned stages three things that are crucial and need to be defined properly are: a)Description of inputs to the system b)description of the processes that are being used and those that will be applied to the system and c) description of the outputs that are desired from the system after processing.These processes though very clear and seemingly succinct to define all have one characteristic , they all are so much absorbing that sometimes it becomes difficult for the analyst as well as the project team that how far should the analysis continue.In other words it entails asking the question:Are we ready to design?.I try and present a model which will try to solve or atleast give the analysts some yardstick by which a solution to this problem can be achieved.I call this model the **AAP(Akshay's analysis paralysis)** model.

The roadmap of the paper is as follows. I motivate the need for analysis in systems analysis and design. I follow by presenting an algorithm which could effectively counter the "analysis paralysis" problem.I then present the AAP model. Ultimately, I conclude.

## 2 Principles

The major principle underlying analysis in systems analysis and design is that the analysis phase serves as a starter or a blueprint for carrying out all the major activities that actually lead t the design and then the implementation of a system that is being developed. Thus this phase is the most crucial in terms of the effect that it has over the entire system. If the problem is not understood or analysed properly then it would lead to a poor design and later on might lead to an entirely wrong implementation. However an over or an underestimation of the problem at hand would result in a catastrophe.So, the major principle underlying analysis effectively means the correct diagnosis and a plan for solving a particular problem.

## 3 Implementation

The question that most of the system administrators/analysts are faced with is when to cross over from the analysis to the design phase? Which effectively means when to stop planning and start designing? A typical systems analyst is responsible for, and acts with the following set of entities.

a) **PEOPLE**, which includes managers, users and other developers.
b) **DATA** including capture, validation, organization, storage and usage.
c) **PROCESSES** both automated and manual.
d) **INTERFACES** both to other systems and applications, as well as to the actual users
e) **GEOGRAPHY** to effectively distribute the data and processes to the people.

[3].Keeping the above mentioned details in mind we realize that a systems analyst has to be involved at all phases of the SDLC. But it is in the first three phases that his/her role is of the utmost significance. In my proposed model the following are the terms/parameters that need to be defined:

**3.1. People Index (PI)**: This actually is a measure that will help the analyst to assess the amount of contribution that the people involved in the analysis stage have done. .There are basically two types of people who are involved in the collection of information:

    **a) Information Gatherers**: Theses are the set of people who are responsible for collecting the information that is needed by them during the information gathering phase. They actually consist of all the analysts, programmers, interviewers etc. who collectively constitute the information gathering team.

**b) Information Sources**: These are the set of people who actually provide the information that is needed by the information gathering team. These may be or may not be a part of the system that is under investigation. Though mostly they consist of the end users i.e. the set of people for whom the system is intended this might not hold true in all the cases.

**PI helps us in assessing the amount of contribution that has come from both these participating entities**. To help set the values of PI for a particular project the following questions can provide a useful estimate:

- How many types of fact finding techniques[4],[5] were followed?
- How many times was the system under study visited?
- What was the composition of the fact finding team?
- What was the experience level of the team?
- To what extent was the literature of the existing system reviewed?
- Did the various questionnaires and feedback forms have some distinguishing entries or was it standard run of the mill stuff?
- How much use of automated tools and technologies made vis a vis manual strategies?
- How much of the information gathered was maintained in proper documented form?
  There are also a number of mathematical formulae like the method of Least Squares [6] that could prove effective, among others.

As the value of PI has been kept between 0 and 1 it means that if the assessment as per the preceding questions falls less than half i.e. 0.5 then the team has not done its work properly and has not been able to gather the requisite information. Which actually implies that the composition of the information gathering team needs rethinking. Hence this is a very crucial factor in deciding whether the analysis can be carried out properly or not.

**3.2 Data gathered (DG)**: This actually helps the analyst in deciding how much data has actually come into his/her possession. The data can be further subdivided into the following two categories:

a) Data that is immediately useful (U): This is the amount of data that is going to come in handy immediately. This essentially means that whatever data has come to the Analyst he/she can, based upon that data take immediate decisions as to the design of the proposed new system. Consider e.g. that a system analyst gets the information that a particular process is not working due to a certain fault which, with a little amount of modification can be easily handled Then the analyst can change or modify the erring process without much of an effort.

b) Data for future use (F): This implies the data that is not immediately relevant or useful but can come in handy at a later stage. Consider e.g. that an analyst who is investigating the Finance Department of a corporate house receives information that 4 Human Resource personnel have left the company in the past one month due to low salary. Though this information seems to be not useful immediately but might come in useful later.
Many techniques have been developed for Data Gathering which may be deduced mathematically.[7],[8].

As the value of U and F has been kept between 0 and 1 it means that if the assessment as per the preceding discussion falls less than half i.e. 0.5 then the actions as per the AAP model algorithm need to be taken.

**3.3 Process Index (PRI)**: This will give the analyst an idea of how much he/she knows of the processes in the existing system. Also based upon this index a strategy of identifying the core problem processes and their rectification can be formulated[9],[10]. The processes can be divided into the following segments:

- **3.3.1** <u>Core Processes</u>: These are the processes that are absolutely essential to the working of the current system. If these processes are removed or develop a fault then the entire system would collapse.
- **3.3.2** <u>Supporting Processes</u>: These are those processes, which though play important roles in the functioning of the system, but they only play a supporting role to the core processes. Thus they indirectly affect the functioning of the system.

The data gathered can actually be compared and analysed by using certain empirical approaches which in association with process identification could produce insightful results[11]Keeping the value of PRI between 0 and 1 means that if this index falls to a value less than 0.5 then the analysis team has not understood the functioning of the processes which actually means that the analysis has to be done again since if processes are not understood then the new system would not prove to be better than the existing one. On the other hand if the value of PRI comes out to be greater than 0.5 and moves towards the value of 1 it means that the processes have been well understood and based upon these have been identified as core or supporting and as such can be put to use in the new system or be totally done away with.

**3.4 Interface Utility (IU)**: Though not strictly a part of the analysis phase this parameter also has to be taken into consideration because finally the effectiveness of the system will be gauged by the users and a poor interface could destroy a well designed system. Speaking in terms of conventional systems development the development of user interface falls into the design phase. However in the analysis phase the analyst can look at the kinds of outputs that a user interface is expected to provide and keeping this in mind can frame a few set of questions which could be used to provide an input to the design phase. Some questions could be like:

- **3.4.1** What are the outputs that are to be expected of the User Interface?
- **3.4.2** What are the processes that could be reflected in the User Interface?
- **3.4.3** Who are the types of users that the interface would cater to?
- **3.4.4** Would a desktop based interface suffice or would it be a web based interface?

The list of questions, though not exhaustive in nature help to identify in general the characteristics that the interface would possess and that would serve as an input to the design phase. Also some help may be derived by following already established techniques such as an Iterative Interface Design[12].If the value of UI tentatively falls short i.e. les than 0.5 then the analysis team needs to rethink and have a meeting with the intended designers of the new system.

**3.5 Geographical Quotient (GQ)**: Again though the issue of distribution crops up in the implementation phase but even in analysis if this factor is taken into account it has a significant bearing on the outcome of the project . **Consider a scenario**:

Let us assume that a team of analysts have successfully implemented an E-governance project in Tamil Nadu(TN). Now the same team is called upon to implement a similar project in Himachal Pradesh(HP). The team based upon its past experience tries a similar strategy but somehow here things do not seem to be going their way. They find that the geographical dissimilarities between the two places have given rise to a number of factors:

- **3.5.1** The preferred mode of UI interaction was English in TN but has to be Hindi in HP
- **3.5.2** The people are resistant to ideas in HP ,which were easily accepted in TN
- **3.5.3** There is a lot of red tapism in HP which was missing in TN
- **3.5.4** The hilly nature of the terrain in HP is making information gathering difficult vis a vis TN.

There could be many more hindrances that the team could have identified. Thus the geographical dissimilarities introduce a wide variety of other issue which could range from social to cultural to economical to anything else. There are a wide variety of techniques that have been mathematically tried and used for gathering spatial data and for accurately identifying them.[13],[14],[15].Using these techniques some sort of data can be acquired which could go some way in solving geographical problems for the analyst.However it is clear that if GQ falls to less than 0.5 it means that the solution we are striving for cannot be much generic in nature and cannot be reused.

**All the above mentioned terms are assumed to have a range of values between 0 and 1 where 0 denotes the base value and 1 denotes the peak value.**

So having defined the terms that the model would use lets now define the steps that the model will have which can be defined using the following algorithm(**AAP model**):

1) Check all the data that has been gathered so far. Based upon the two criteria i.e. immediate usefulness and future use divide it into two relevant portions.
2) For each of the two different piles of data so created make a listing of the contribution of the people involved (PI).
3) a) If PI > 0.5 but either U<0.5 or F<0.5 then the data gathered has no relevance. Discard and begin analysis afresh.
   b) If PI<0.5 but U>0.5 then data is handy. Check for future usefulness.
   c) If PI<0.5 but U>0.5 and F>0.5 then you have data that is both useful and handy for future.
However the team needs reworking.
   d) If PI>0.5 but U>0.5 and F>0.5 then you have data that is both useful and handy for future.
Move to the next stage.
4) Out of the data gathered so far again divide the data into two piles. Again check for the immediate usefulness and future use. However this time these parameters would be defined relative to only one factor i.e. Process Index (PRI).
5) If PRI<0.5 then analysis needs to be done afresh. Back to step 1.
6) If PRI>0.5 but <1 then analysis needs to be continued but only a greater understanding of the processes needs to be had and shouldn't start from scratch.
7) If PRI=1 then we can move to the next step.
8) Check for IU. If IU<0.5 then rethink. Check also if PI<0.5.If both conditions hold then involve more people. Else move to the next step.
9) Check if GQ<0.5.Also check if either PI<0.5 or PRI<0.5.If yes then maybe GQ is affecting both PI and PRI to some extent. Try and find alternatives.
10) If GQ>0.5 then we are ready to move into the design phase.

As is clear from the above algorithm we can actually try and assess our progress until we have reached a desired level.

**4  Diagrammatic Representation of the AAP Model**

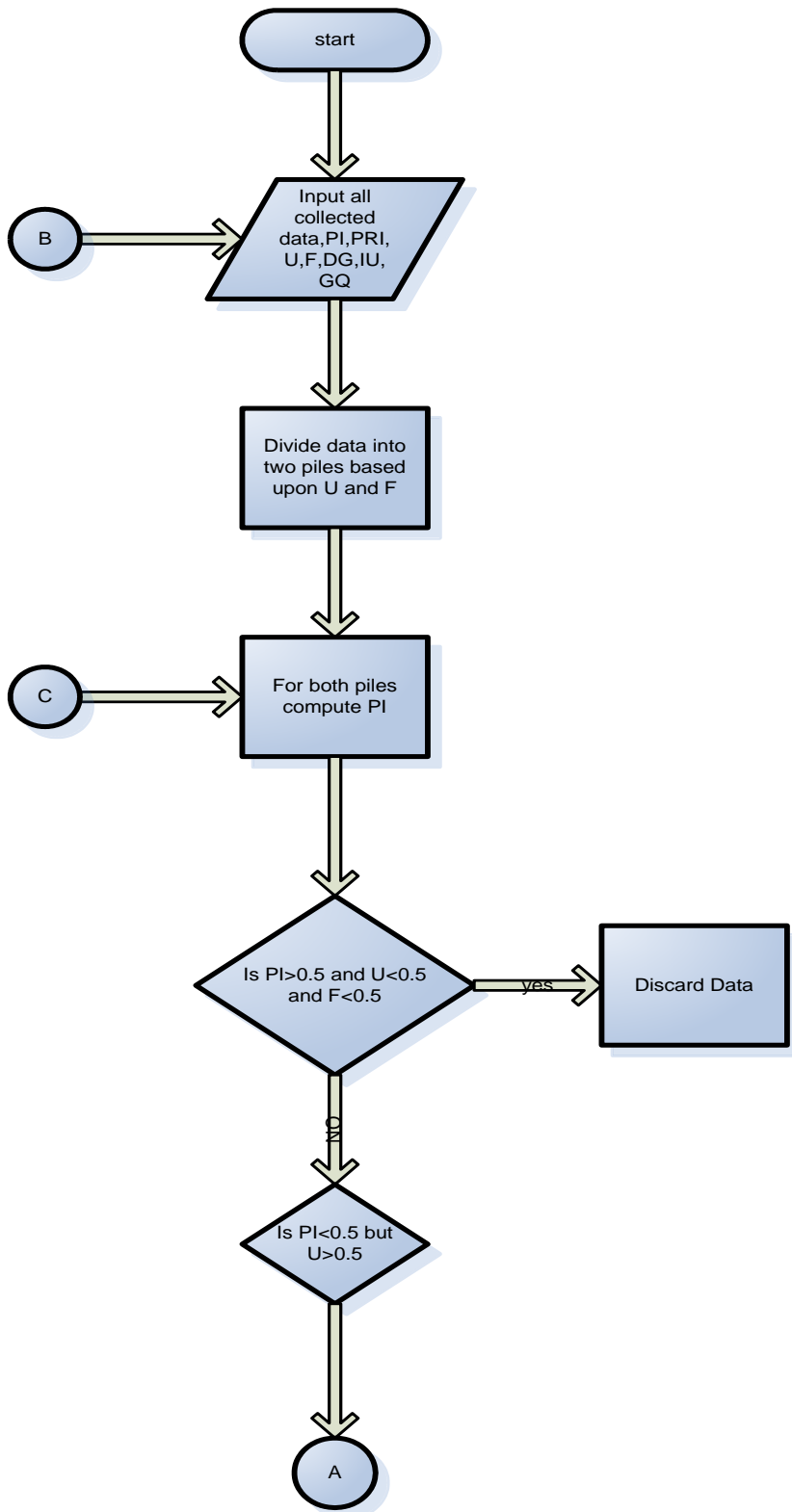

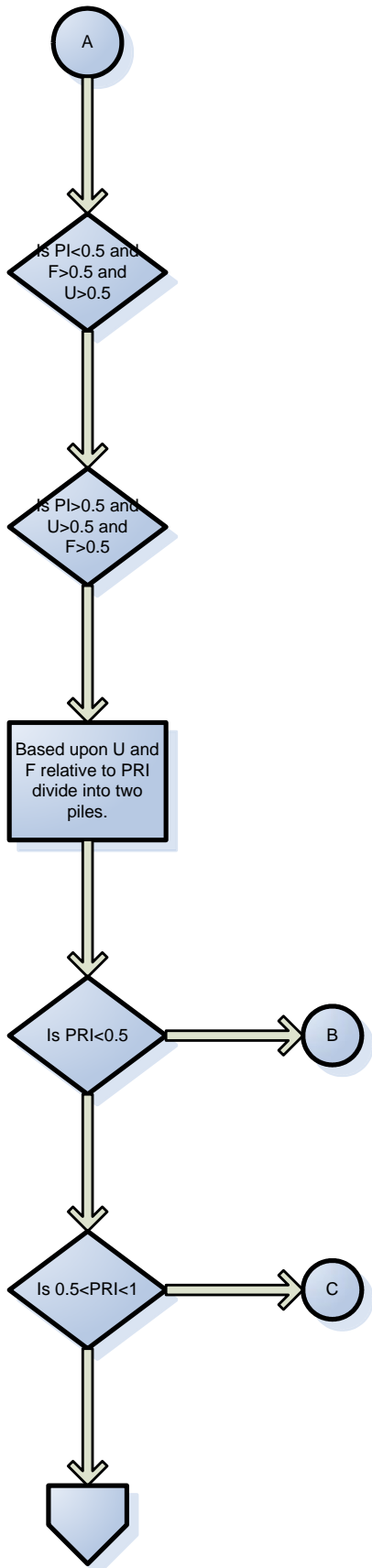

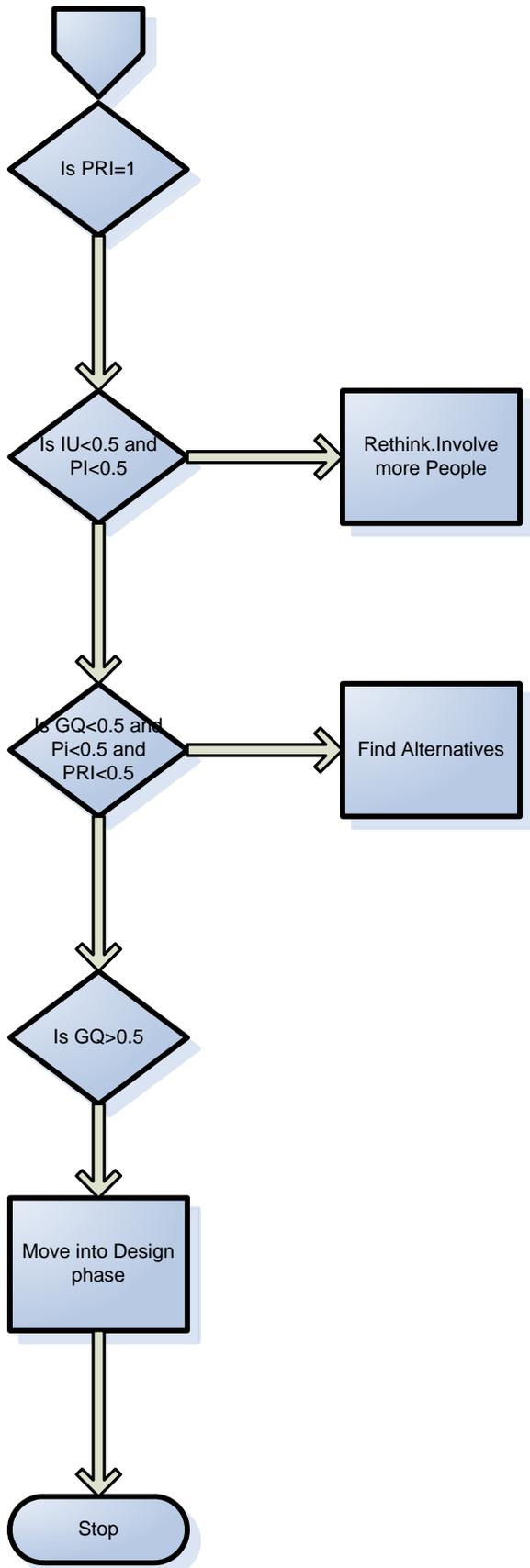

**Figure 1**

## 5 Conclusion

All the above mentioned terms are assumed to have a range of values between 0 and 1 where 0 denotes the base value and 1 denotes the peak value. The question that crops up in all the above cases is how to decide the values for each of the parameters?

This is partially based upon the discretion of the systems analyst or whoever is in charge of the analysis team. It is him/her/they who will decide what range of values they give for each of the factors because the criteria for these would change from project to project.However the margin for dicretion would be very snmall as numerous mathematical and statistical tools exist to aid the analyst as pointed out by the preceding discussion.The data collected by the numerous techniques referenced to above could be normalized by mathematical tools to fall between 0 and 1. The AAP model and algorithm serves as a generic guideline and provides a yardstick to the analyst to gauge and assess the answer of when and where to stop analyzing. In the preceding discussion, however we have almost always advanced to the next step if we get values >0.5.This is so because in trying to reach the value of 1, we actually fall into the trap of "Analysis Paralysis"